\documentclass[reprint, aps, pra, floatfix, superscriptaddress]{revtex4-1} 
\usepackage{graphicx}
\usepackage{epstopdf}
\usepackage{bm, amsmath, amssymb}
\usepackage{float}
\usepackage{placeins}
\usepackage{color}
\usepackage{array}
\usepackage{siunitx}
\usepackage{array}
\usepackage{makecell}
\usepackage[colorlinks=true, allcolors=blue]{hyperref}
\usepackage{multirow}
\usepackage{dblfloatfix}

\newcommand{\mK}{\ \mathrm{mK}}

\newcommand{\mm}{\ \mathrm{mm}}
\newcommand{\um}{\ \si{\um}}

\newcommand{\us}{\ \mathrm{\mu s}}

\newcommand{\kHz}{\ \mathrm{kHz}}
\newcommand{\MHz}{\ \mathrm{MHz}}
\newcommand{\GHz}{\ \mathrm{GHz}}
\newcommand{\dB}[0]{\ \mathrm{dB}}

\begin{document}

\title{Cavity Attenuators for Superconducting Qubits}

\author{Z. Wang}
\email{zhixin.wang@yale.edu}
\affiliation{Department of Applied Physics, Yale University, New Haven, CT 06520, USA}

\author{S. Shankar}
\affiliation{Department of Applied Physics, Yale University, New Haven, CT 06520, USA}

\author{Z. K. Minev}
\affiliation{Department of Applied Physics, Yale University, New Haven, CT 06520, USA}

\author{P. Campagne-Ibarcq}
\affiliation{Department of Applied Physics, Yale University, New Haven, CT 06520, USA}

\author{A. Narla}
\affiliation{Department of Applied Physics, Yale University, New Haven, CT 06520, USA}

\author{M. H. Devoret}
\email{michel.devoret@yale.edu}
\affiliation{Department of Applied Physics, Yale University, New Haven, CT 06520, USA}

\date{\today}

\begin{abstract}
Dephasing induced by residual thermal photons in the readout resonator is a leading factor limiting the coherence times of qubits in the circuit QED architecture. This residual thermal population, of the order of $10^{-1}$--$10^{-3}$, is suspected to arise from noise impinging on the resonator from its input and output ports. To address this problem, we designed and tested a new type of band-pass microwave attenuator that consists of a dissipative cavity well thermalized to the mixing chamber stage of a dilution refrigerator. By adding such a cavity attenuator inline with a 3D superconducting cavity housing a transmon qubit, we have reproducibly measured increased qubit coherence times. At base temperature, through Hahn echo experiment, we measured $T_{2\mathrm{e}}/2T_1 = 1.0\,({+0.0}/{-0.1})$ for two qubits over multiple cooldowns. Through noise-induced dephasing measurement, we obtained an upper bound $2\times 10^{-4}$ on the residual photon population in the fundamental mode of the readout cavity, which to our knowledge is the lowest value reported so far. These results validate an effective method for protecting qubits against photon noise, which can be developed into a standard technology for quantum circuit experiments.

\end{abstract}

\maketitle

\section{Introduction}

The past two decades have witnessed an exponential increase in the coherence times of superconducting qubits \cite{Devoret13}. Major breakthroughs include biasing a charge qubit at the charge-noise-insensitive point \cite{Vion02}, shunting the Josephson junction with a capacitance to suppress noise in charge \cite{Cottet02, Koch07, Schreier08} and flux qubits \cite{You07, Steffen10}, and shunting the junction with a superinductance to eliminate offset charge noise in a fluxonium qubit \cite{Manucharyan09}. Moreover, in the circuit QED architecture \cite{Blais04, Wallraff04}, it has been well understood that the readout cavity modifies the electromagnetic environment of the qubit and can thus affect qubit coherence \cite{Esteve86, Houck08}. In recent years, for transmon qubits embedded in 3D microwave cavities \cite{Paik11}, energy relaxation times $T_1>100\us$ have been frequently observed \cite{Riste13, Dial16, Narla16, Minev18}. However, in these experiments, transmon coherence times $T_2$ are much shorter than $2 T_1$, indicating qubit coherence is predominantly limited by pure dephasing. An outstanding question in this field is whether $T_2 \approx 2T_1$ can be reproducibly obtained in long-lifetime superconducting qubits.

One of the main dephasing channels that limit $T_2$ is the residual thermal photon population in the readout cavity \cite{Bertet05a, Bertet05b, Clerk07, Sears12, Yan16, Yeh17, Yan18}. In the dispersive coupling regime, due to the AC Stark effect, fluctuations in the thermal photon number of the cavity cause random shifts of the qubit transition frequency and thus contribute to qubit dephasing \cite{Gambetta06}. In the limit of average thermal photon number $\bar{n}_\mathrm{th}\ll 1$, the induced dephasing rate is proportional to $\bar{n}_\mathrm{th}$ \cite{Clerk07}:
\begin{align}\label{dephase}
  \Gamma_\phi^{\mathrm{th}} = \frac{\bar{n}_\mathrm{th}\kappa \chi^2}{\kappa^2+\chi^2},
\end{align}
where $\kappa$ is the cavity linewidth, and $\chi$ is the dispersive shift of the qubit frequency per cavity photon. Theoretically, for a $7.5 \GHz$ cavity mode at $20 \mK$, $\bar{n}_\mathrm{th}$ is expected to be on the order of $10^{-8}$. However, the $\bar{n}_\mathrm{th}$'s estimated from measurements of $\Gamma_\phi^{\mathrm{th}}$ in recent experiments range from $6\times 10^{-4}$ to 0.15, corresponding to effective mode temperatures between $55 \mK$ and $140 \mK$ \cite{Suri13, Rigetti12, Yan16, Yeh17, Goetz17, Yan18}. Understanding the origin of excess thermal photons and reducing $\bar{n}_\mathrm{th}$ are therefore crucial to enhancing the qubit coherence times and reliably achieving $T_2 \approx 2T_1$.

One source of the excess $\bar{n}_\mathrm{th}$ is the coupling of the cavity mode to the input and output ports. This coupling opens a channel for auxiliary components in the microwave wiring to affect the temperature of the cavity mode. Examples of such components are commercial cryogenic attenuators, filters, isolators, etc. These components and the teflon insulator in the coaxial cables are difficult to thermalize to the mixing chamber stage of the dilution fridge. Attenuators are particularly important in this regard since they are the dominant dissipation sources in the wiring of a cryostat and ideally form the bath that thermalizes the readout cavity modes. Insulators inside commercial cryogenic attenuators have poor thermal conductivities at low temperatures as do their stainless steel packages. Lately, thermal anchoring of attenuators has been improved by replacing these materials with better thermal conductors and redesigning the circuit layout \cite{Yeh17}. Nevertheless, a fundamental restriction in performance arises from the lumped element thin-film resistive network that these attenuators are made of.

\begin{figure} [t]
 \includegraphics[angle = 0, width = \columnwidth]{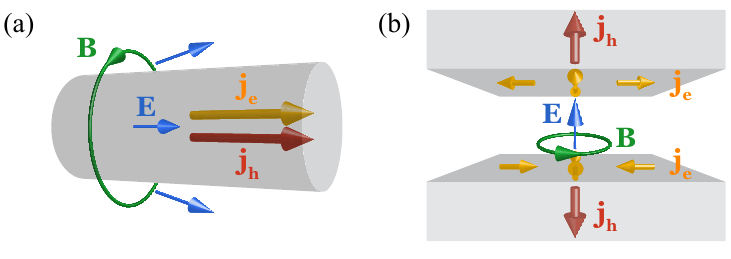}
 \caption{Comparison of the electromagnetic fields outside and currents inside (a) a lumped element resistor and (b) a dissipative waveguide section (TE mode). Blue and green arrows indicate electric and magnetic field lines, while yellow and red arrows indicate electric current $\mathbf{j}_\mathrm{e}$ and heat current $\mathbf{j}_\mathrm{h}$, respectively.
 }
 \label{fig1}
\end{figure}

An explanation of this challenge and its solution are shown in Fig.~\ref{fig1}. In a lumped element resistor, the electric and heat currents are parallel to each other [see Fig.~\ref{fig1}(a)]. Given a certain electrical resistance, the order of magnitude of its thermal resistance due to electronic degrees of freedom is subject to the constraint imposed by Wiedemann--Franz law \cite{Ashcroft}. For instance, a $50\ \si{\ohm}$ resistor at $20 \mK$ should have an electronic thermal resistance on the order of $ 10^{8}\ \mathrm{mK/\si{\micro}W}$. Therefore, microwave attenuators at low temperatures are primarily thermalized through electron--phonon interaction and phonon transport \cite{Roukes85, Wellstood94}. However, this phonon mechanism suffers from a bottleneck since very few phonons are present at millikelvin temperatures. Numerical simulations show that under microwatt input power, the temperature difference inside the resistive network of the attenuator can easily reach $100 \mK$ \cite{Yeh17}. While it may be reduced by proper choice of materials and circuit layout, thermalizing all materials to 20 mK may be fundamentally impossible, since some thermal gradient is necessary to produce a sufficient phonon heat current.

To overcome this fundamental challenge, we would prefer to use electrons in a Fermi degenerate system to conduct heat, since electronic excitations are always present at low temperatures. We thus consider an alternative dissipation source for microwave radiation---the normal metal walls of a waveguide section. As shown in Fig.~\ref{fig1}(b), in a dissipative waveguide, electric current mainly flows within the skin depth of the wall, while heat is conducted by electrons into the bulk metal, perpendicular to the electric current. In principle, such a distributed structure, if made of a good thermal conductor, can realize an improved cold black body radiation environment at microwave frequencies, which is the goal of our work.

We present in this paper a new type of narrowband microwave attenuator based on a cavity realized by a section of dissipative waveguide, which aims at reducing the residual thermal photon population in circuit QED systems. The normal metal cavity is seamless and machined by conventional techniques without any material growth or microfabrication processes. By coupling it to a transmon--cavity system, we measured enhanced coherence times on two qubits in multiple cooldowns, and obtained record-breaking pure dephasing time $T_\phi$ and $\bar{n}_\mathrm{th}$ at base temperature, leaving qubit coherence limited by energy relaxation. Because of its simple design and reliable performance, this cold cavity attenuator will provide a useful addition to the state-of-the-art quantum circuit toolbox.


\section{Cavity attenuator design}

Most commercial cryogenic attenuators have more than $10 \GHz$ bandwidth. However, microwave pulses for qubit control and measurement are centered only around certain frequencies, and thus only narrowband attenuation is required. This forms the basis for the idea of attenuating quantum signals with a dissipative cavity. For instance, to protect a readout cavity at frequency $f_\mathrm{r}$ with $\sim 1 \MHz$ linewidth from excess photon noise, we would like to supplement commercial attenuators with a cavity attenuator. Such a cavity attenuator should: (1) be centered around $f_\mathrm{r}$, (2) provide 10--20 dB attenuation on resonance, and (3) have 10--50 MHz bandwidth to cover the linewidth of the readout cavity. In addition, it should be made of a low-temperature-compatible normal metal, such as brass or, even better, oxygen-free high thermal conductivity (OFHC) copper.

Designing a microwave cavity that can be well thermalized to millikelvin temperatures and has a 10--50 MHz internal dissipation rate (internal quality factor $Q_\mathrm{i} \sim 500$) is a challenging task, since good thermal conductors also have low electrical conductor loss. Given the resistivity of the material, the quality factor of a 3D cavity resonator is approximately inversely proportional to its surface-to-volume ratio \cite{Pozar}. An order-of-magnitude estimate shows that the smallest dimension of our brass (OHFC copper) dissipative cavity must be in the submillimeter (sub-0.1-mm) regime. Furthermore, the on-resonance power transmission of a two-port resonator is $ 4 \kappa_{\mathrm{c}1} \kappa_{\mathrm{c}2} / (\kappa_\mathrm{i} + \kappa_{\mathrm{c}1} + \kappa_{\mathrm{c}2})^2$, where $\kappa_\mathrm{i}$ and $\kappa_{\mathrm{c}1,2}$ are the internal dissipation and external coupling rates. 10--20 dB attenuation requires $\kappa_\mathrm{i}\sim 10 \kappa_{\mathrm{c}1,2} $---the cavity being undercoupled to the couplers.

\begin{figure} [htb]
 \includegraphics[angle = 0, width = \columnwidth]{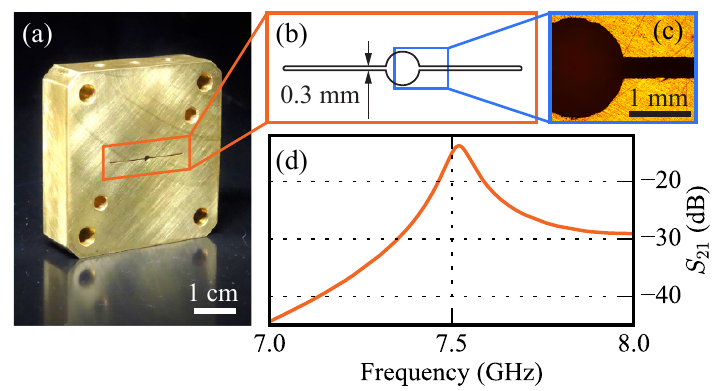}
 \caption{Brass cavity attenuator. (a) Photography of the device. (b) Cross-section drawing of the thin seamless cavity. (c) Zoomed-in optical micrograph. (d) Cavity transmission at room temperature.
 }
 \label{fig2}
\end{figure}

Fig.~\ref{fig2} shows a physical realization satisfying all these requirements. It is a single-piece brass block fabricated using wire electrical discharge machining (EDM) \cite{Ho04, EDM}. Such a design eliminates seams and therefore their associated loss \cite{Brecht17}. Its external dimensions match commercial WR-102 waveguides. In addition, it can be directly thermalized to the mixing chamber of a dilution refrigerator through copper braids. As sketched in Fig.~\ref{fig2}(b), the cavity is rectangular, apart from a cylindrical hole in the middle for initializing wire EDM cutting. Due to the open boundary condition at both coupling ports, the electromagnetic field in the fundamental mode is non-uniform only in the longest dimension that is $22 \mm$ in this device and sets the mode frequency. The $0.3 \mm$ gap sets the internal dissipation rate. We characterized this cavity at room temperature with a calibrated network analyzer. The waveguide couplers on both sides have $\kappa_\mathrm{c} \approx \kappa_\mathrm{i} /10$. As seen in Fig.~\ref{fig2}(d), the transmission peak centers at $7.52 \GHz$ with $14 \dB$ insertion loss. Combined with reflection measurements, we extracted $\kappa_\mathrm{i} /2\pi = 54 \MHz$.

Based on this design, we fabricated cavity attenuators with different materials and gap sizes. Parameters of three representative devices are listed in Table~\ref{tab1}. For cryogenic measurements, the temperature of the attenuator block reached 15~mK as verified by $^{60}\mathrm{Co}$ nuclear orientation thermometry. As seen in Table~\ref{tab1}, cavity frequencies increase due to thermal contraction at low temperature. Meanwhile, phonon vibrations freeze out, which increases the metal conductivities and reduces the cavity dissipation rates. For brass, a copper-zinc alloy, $\kappa_\mathrm{i}$ is reduced by $\sim 20\%$. For OFHC copper, $\kappa_\mathrm{i}$ is reduced by around a factor of three, indicating a ten-fold increase in electrical conductivity. Note that this value is smaller than the DC residual-resistivity ratio (RRR) of OFHC copper because of the anomalous skin effect, which appears when the mean free path of the metal is longer than the wavelength of the RF probe signal \cite{Pippard47, Chambers50, Pippard54}. The dissipation rates of copper cavity attenuators are limited by the smallest diameter of the EDM cutting wires that our machine shop has access to.

\begin{table} [tbh]
\centering
\begin{tabular}{ >{\centering\arraybackslash}m{1.75cm} |>{\centering\arraybackslash}m{2.0cm} |>{\centering\arraybackslash}m{1.0cm} >{\centering\arraybackslash}m{1.0cm} >{\centering\arraybackslash}m{1.0cm}}
 \hline
 \hline
 \multicolumn{2}{c|}{Material} & Brass & Cu & Cu \\
 \hline
 \multicolumn{2}{c|}{Gap ($\si{\um}$)} & 300 & 75 & 125 \\
 \hline
 \multirow{2}{*}{$T = 296\ \mathrm{K}$} & $f_\mathrm{a}$ (GHz) & 7.52 & 7.68 & 7.68 \\
 & $\kappa_\mathrm{i}/2\pi$ (MHz) & 54 & 69 & 62 \\
 \hline
 \multirow{2}{*}{$T = 15\ \mathrm{mK}$} & $f_\mathrm{a}$ (GHz) & 7.67 & 7.79 & 7.75 \\
 & $\kappa_\mathrm{i}/2\pi$ (MHz) & 44 & 19 & 24 \\
 \hline
 \hline
\end{tabular}
\caption{Resonant frequency $f_\mathrm{a}$ and internal dissipation rate $\kappa_\mathrm{i}$ of three cavity attenuators measured at room temperature and 15 mK.}
\label{tab1}
\end{table}


\section{Qubit coherence enhancement} \label{coherence}

\begin{figure} [htb]
 \includegraphics[angle = 0, width = \columnwidth]{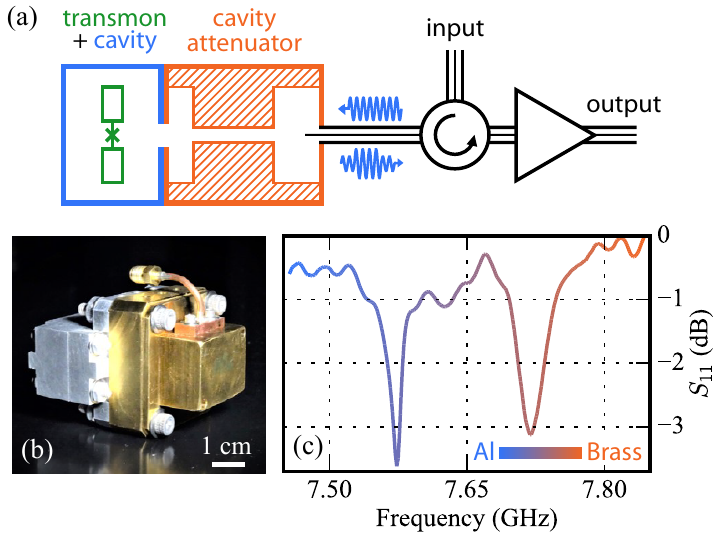}
 \caption{Test of attenuator with transmon--cavity system. (a) Measurement setup anchored to the base stage of a dilution refrigerator. A transmon qubit (green) is dispersively coupled to the fundamental mode of an aluminum readout cavity (blue), which is itself coupled to a cavity attenuator (orange) through an aperture. Between the two hybridized modes, the one with larger participation in the aluminum cavity is used to readout the qubit in reflection. (b) Photograph of the coupled cavities. The transmon is held inside the aluminum cavity (left). The cavity attenuator (middle) is coupled to the microwave lines through a coaxial cable coupler (right). (c) Cavity reflection at $13 \mK$ showing the hybridized modes. Colorbar indicates the participations of the mode in the aluminum (blue) and the brass (orange) cavities.
 }
 \label{fig3}
\end{figure}

We tested the performance of these cavity attenuators by coupling them to a circuit QED system and measuring the qubit coherence properties between 13 mK and 120 mK. The experimental setup is shown in Fig.~\ref{fig3}. A cavity attenuator is connected to the sole coupling port of the aluminum cavity housing a transmon qubit. Microwave measurement of the cavity can be performed in reflection. This configuration provides maximum protection against incident photon noise---the transmon is not directly exposed to any excess radiation coming from the transmission lines. Consequently, thermal-photon-induced qubit dephasing can be studied in detail. However, this configuration also degrades the measurement signal-to-noise ratio because the readout signal has been attenuated before being amplified by the output chain. This problem could be solved by \textit{in situ} quantum limited amplification, which is discussed in Sec. \ref{diss}. We note that one of the ideas behind our experiment---introducing a cold dissipation source in the path of quantum signals---is similar to that of a previous work \cite{Rigetti12}. Alternatively, the qubit dephasing can also be suppressed by reducing the coupling rate of the readout cavity to the output line \cite{Sears12}. Nevertheless, compared to these strategies, our two-cavity modular approach provides more flexibility in experimental design. In addition, the cavity attenuator filters out off-resonance radiation and thus further suppresses the radiative decay of the qubit \cite{Esteve86, Houck08}, acting as an effective Purcell filter \cite{Reed10}.

We measured the 0.3 mm-gap brass cavity attenuator coupled to a transmon--aluminum readout cavity system. The reflected signal off the combined system, measured using a network analyzer [see Fig.~\ref{fig3}(c)], shows two hybridized modes centered at $7.573 \GHz$ and  $7.719 \GHz$. Using the measured trace, we estimated that the mode centered at $7.573 \GHz$ participates 79\% in the aluminum cavity and 21\% in the brass cavity, while the mode centered at $7.719 \GHz$ participates 21\% in the aluminum cavity and 79\% in the brass cavity. We used the mode at $7.573 \GHz$ to readout the qubit since it participates more in the aluminum cavity and thus has a larger dispersive shift. The ratio of internal dissipation to external coupling for this readout mode is estimated to be 6:1. Denoting the average photon population of the internal and external baths by $\bar{n}_\mathrm{i}$ and $\bar{n}_\mathrm{c}$, we have $\bar{n}_\mathrm{i} \ll \bar{n}_\mathrm{c}$ since the brass cavity is in thermal equilibrium with the mixing chamber. Therefore, we obtain the residual thermal population of the readout mode to be $\bar{n}_\mathrm{th} \approx \bar{n}_\mathrm{c} / 7$, indicating 85\% of residual photons are dissipated in the cold cavity attenuator.

\begin{table} [tbh]
\centering
\begin{tabular}{ >{\centering\arraybackslash}m{2.0cm} | >{\centering\arraybackslash}m{1.5cm} >{\centering\arraybackslash}m{1.75cm} }
 \hline
 \hline
 Transmon & $f_{ge}$ (GHz) & $\alpha$ (GHz) \\
 \hline
 A & 4.75 & 0.25\\
 B & 5.09 & 0.25\\
 \hline
 \hline
\end{tabular}
\caption{$f_{ge}$ and $\alpha = f_{ge} - f_{ef}$ (anharmonicity) of the two transmon qubits in this experiment.}
\label{tab2}
\end{table}

\begin{figure} [htb]
 \includegraphics[angle = 0, width = \columnwidth]{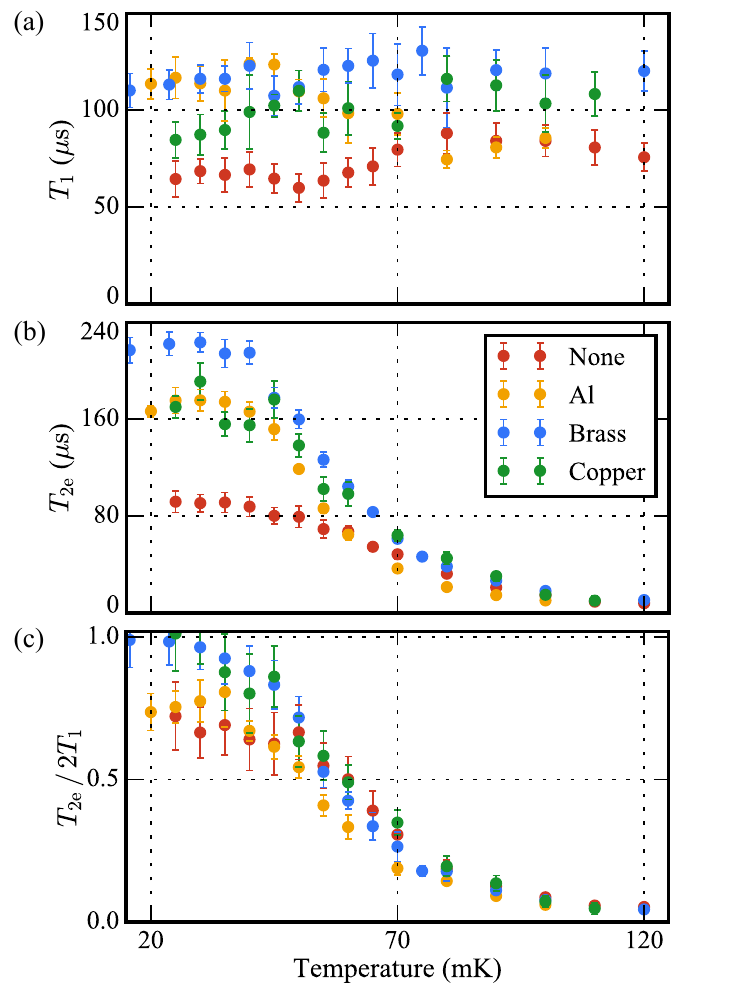}
 \caption{Qubit coherence data. (a) $T_1$, (b) Hahn echo $T_{2\mathrm{e}}$, and (c) $T_{2\mathrm{e}}/2T_1$ versus temperature with and without cavity attenuators. All the data are collected on transmon A. Error bars include both the measurement imprecision and the fluctuation of $T_1$ over the one hour data acquisition time. Red circles: no attenuator. Yellow circles: aluminum cavity attenuator with $75 \um$ gap. Blue circles: brass cavity attenuator with $300 \um$ gap. Green circles: copper cavity attenuator with $75 \um$ gap.
 }
 \label{fig4}
\end{figure}

$T_1$ and Hahn echo $T_{2\mathrm{e}}$ were measured as a function of temperature for two transmon--cavity systems, labeled A and B, coupled with brass and copper cavity attenuators as well as without an attenuator. As a control experiment, we also measured an aluminum filter with identical dimensions to the copper attenuator that however provides no attenuation on resonance. The transmon frequencies and anharmonicities are listed in Table~\ref{tab2}. Data taken on transmon A are shown in Figs.~\ref{fig4}(a) and \ref{fig4}(b), with experimental conditions summarized in Table~\ref{tab3}. Each $T_1$ and $T_{2\mathrm{e}}$ data point is the average of ten measurements performed over the course of around one hour.

From these results we can infer that cavity attenuators dissipate excess photons in the readout mode and suppress photon-induced qubit dephasing. We see in Fig.~\ref{fig4}(b) that $T_{2\mathrm{e}}$ for transmon A at base temperature is improved by more than a factor of two with cavity attenuators and can exceed 220 \si{\us}. Meanwhile, Ramsey $T_{2\mathrm{R}}$ (not shown) was also improved from 28--35 \si{\us} without a cavity attenuator to 41--43 \si{\us} with the brass attenuator and 35--40 \si{\us} with the copper attenuator. The difference between $T_{2\mathrm{R}}$ and $T_{2\mathrm{e}}$ indicates that low-frequency noise in our measurement setup is causing qubit dephasing. However, since the dephasing due to residual thermal photons cannot be filtered out by a single echo pulse \cite{Sears12}, we use $T_{2\mathrm{e}}$ to inform us about the effect of the cavity attenuator on the residual thermal photon population.


An important figure of merit to quantify qubit dephasing is the dimensionless value $T_{2\mathrm{e}} / 2T_1 = T_\phi / (T_\phi + 2 T_1)$, which is close to its unity upper limit when the dephasing time satisfies $T_\phi \gg T_1$. As shown in Fig. \ref{fig4}(c), at base temperature we measured $T_{2\mathrm{e}} / 2T_1 = 0.98\, ({+0.02}/{-0.08})$ with the brass attenuator. This ratio is $1.00\, ({+0.00}/{-0.12})$ with the copper attenuator. In either experiment, the average $T_\phi$ is close to 10 \si{\ms}, much longer than $T_1$, indicating qubit coherence is limited by relaxation rather than pure dephasing. If we attribute all the qubit dephasing to the residual thermal photon population in the fundamental mode of the readout cavity, according to Eq. (\ref{dephase}), the upper bound of $\bar{n}_\mathrm{th}$ is estimated to be on the order of $10^{-4}$, corresponding to an effective mode temperature of $T_\mathrm{eff} \leq 40$--$45 \mK$. As a comparison, in the absence of cavity attenuators, transmon A in the same measurement setup has $T_\phi \approx 0.3\ \si{\ms}$ and $T_{2\mathrm{e}} / 2T_1 = 0.72\pm 0.12$, indicating $\bar{n}_\mathrm{th} \leq 4 \times 10^{-3}$ and $T_\mathrm{eff}\leq 65 \mK$.

To verify the efficacy of these cavity attenuators, we further performed two control experiments. First, in the same geometry as the 75-\si{\um}-gap copper attenuator, we machined an aluminum cavity. It becomes a lossless cavity filter below 1 K and thus should leave the thermal photon population of the readout mode unchanged. By coupling it to transmon A and performing the same temperature-dependent measurements, we acquired the yellow circles in Fig. \ref{fig4}. At 25 mK, we measured $T_{2\mathrm{e}} / 2T_1 = 0.75\pm 0.06$, indicating $\bar{n}_\mathrm{th} \leq 1 \times 10^{-3}$ and $T_\mathrm{eff}\leq55 \mK$, which are between the no-attenuator and brass/copper-attenuator results. Therefore, we conclude that the enhancement of qubit coherence in previous experiments is not only due to lossless filtering. Dissipation is necessary for reducing high-frequency dephasing noise caused by residual thermal photons in the readout mode.

As a second control, we repeated the copper-attenuator experiment but added to the aluminum readout cavity a copper coaxial cable coupler with coupling rate to the readout mode $\sim 5 \kHz \ll \kappa_\mathrm{c}, \kappa_\mathrm{i}$. This input line with 70 dB cold attenuation was terminated by a 50 \si{\ohm} load at room temperature. At 25 mK, we observed $T_1=100\pm 8\ \si{\us}$, $T_{2\mathrm{e}} = 171\pm 10\ \si{\us}$, and obtained $T_{2\mathrm{e}} / 2T_1 = 0.86\pm 0.09$, $\bar{n}_\mathrm{th} \leq 1 \times 10^{-3}$, corresponding to $T_\mathrm{eff}\leq55 \mK$. We conclude that even a weakly coupled port has nonegligible contribution to qubit dephasing if it is not directly thermalized to the mixing chamber. Based on this and other experiments in our lab, we suspect in particular that the teflon in the coaxial cable is a source of excess photons. Therefore, to achieve the best qubit coherence, ideally every coupling port on the readout cavity should be properly protected by a cold cavity attenuator.

\begin{figure} [htb]
 \includegraphics[angle = 0, width = \columnwidth]{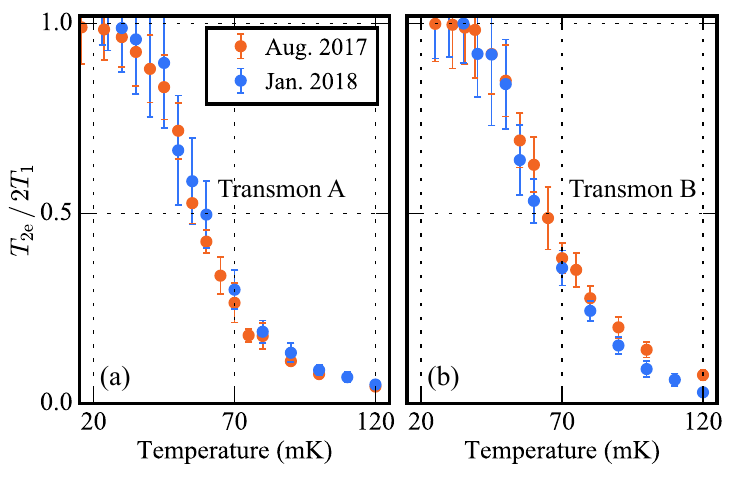}
 \caption{Consistency of $T_{2\mathrm{e}}/2T_1$ across multiple samples and cooldowns when the qubits are protected by $300 \um$ brass cavity attenuators. For transmon B, $T_1 \approx 50\pm 10\ \si{\us}$ below 100 mK.
 }
 \label{fig5}
\end{figure}

We tested the consistency of the performance of cavity attenuators by conducting experiments on the two transmons A and B in multiple cooldowns. Plotted in Fig.~\ref{fig5} are results with 300-\si{\um}-gap brass attenuators. These two cooldowns were separated by five months during which time the attenuators were removed from the setups. For both qubits, temperature-dependent measurements showed good reproducibility on $T_{2\mathrm{e}}/2T_1$, which is close to 1 in average and larger than 0.9 within one standard deviation at base temperature. We believe that reliable performance of these attenuators arises from their well-understood materials, structure, and fabrication process.

\section{Thermal photon population measurement}

\begin{figure} [t]
 \includegraphics[angle = 0, width = \columnwidth]{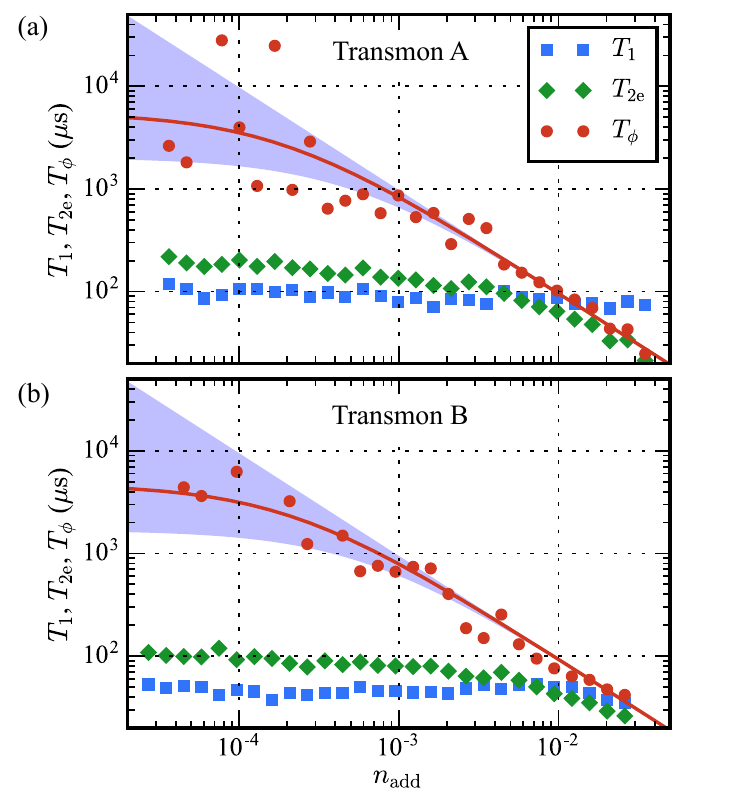}
 \caption{Noise-induced dephasing measurement of $\bar{n}_\mathrm{th}$ in the readout mode. (a) Transmon A protected by a $75 \um$ copper cavity attenuator. (b) Transmon B protected by an $125 \um$ copper cavity attenuator. In each subfigure, $T_1$ (blue squares) and $T_{2\mathrm{e}}$ (green diamonds) are plotted versus added thermal photon number $n_\mathrm{add}$. The extracted $T_\phi$ (red circles) are shown together with a fit (solid curve). Purple shadow represents the fitting error within one standard deviation.}
 \label{fig6}
\end{figure}

Precisely measuring $T_\phi$ and $\bar{n}_\mathrm{th}$ becomes a challenging task when $T_2 \approx 2T_1$, as the fluctuation of $T_1$ over time causes the error bar of $\Gamma_\phi = 1/T_\phi$ to exceed $\Gamma_\phi$ itself. To determine residual $\bar{n}_\mathrm{th}$ at base temperature with higher accuracy, we performed noise-induced dephasing measurement on both transmons. Adopting a method similar to Refs.~\cite{Yan16, Yan18}, we amplitude-modulated broadband white noise (0--80 MHz) onto a continuous-wave microwave signal at the hybridized readout frequency. The total photon population in the readout mode becomes $n_\mathrm{tot} = n_\mathrm{add} + n_\mathrm{th}$, in which $n_\mathrm{add}$ is proportional to the output power of the noise generators. By measuring $n_\mathrm{tot}$ as a function of the noise power (see Fig.~\ref{fig6}), we can extract $n_\mathrm{th}$ by linear regression. Finally, we obtained $n_\mathrm{th} = 2\,(+3/{-2})\times10^{-4}$ for transmon A and $n_\mathrm{th} = 2\,(+4/{-2})\times10^{-4}$ for transmon B when they are protected by copper cavity attenuators, corresponding to $T_\mathrm{eff} \leq 44\mK$. These results are consistent with the order-of-magnitude estimate in Sec. \ref{coherence}, and lower than the values in two other recent reports \cite{Yeh17,Yan18} that also aimed at reducing $\bar{n}_\mathrm{th}$ in circuit QED systems.

\section{Discussion} \label{diss}

The results of our experiments have two important implications: First, cavity attenuators can reproducibly reduce $\bar{n}_\mathrm{th}$ of the readout cavity by an order of magnitude and extend $T_{2\mathrm{e}}$ close to the $2T_1$ limit. Second, the improvement of qubit coherence by cavity attenuators is impacted if the readout cavity has any direct coupling to the input and output lines, even if the coupling is very weak. Our experiment suggests that excess thermal photons inevitably come from the commercial microwave components at base temperature. Consequently, cavity attenuators should be employed as a standard device to create a cold black body radiation environment for superconducting quantum circuits.

Two problems need to be solved before cavity attenuators can have broader applications in circuit QED experiments: First, in the experiments reported in this paper, qubit control and readout tones share the same coupling aperture protected by a cavity attenuator centered near the readout frequency. As a result, qubit control pulses are mostly filtered out, which slows down the qubit state manipulation. This can be avoided by separating the qubit control and readout ports and protecting them with cavity attenuators centered at different frequencies. More desirably, it is possible to design a multi-pole dissipative filter whose bandwidth covers a few qubits simultaneously. Second, in the current measurement setup, while dissipating excess thermal photons, the cavity attenuator also attenuates the readout signal. Consequently, the measurement signal-to-noise ratio is not sufficient for high-fidelity single-shot dispersive readout. This problem might be addressed by moving the cavity attenuator to the output port of the quantum-limited amplifier. More compactly, we could also realize \emph{in situ} quantum-limited amplification in the readout cavity using Josephson junctions or ``SNAILs'' \cite{Narla14, Frattini17}. Combined together, these solutions would bypass the conflict between the isolation and controllability of superconducting qubits, and therefore significantly enhance the qubit coherence times while retaining the benefits of strong qubit--photon interactions in artificial quantum circuits.

\section*{Acknowledgements}
We acknowledge insightful discussions with I. Tsioutsios and U. Vool.
Facilities use was supported by the Yale Institute for Nanoscience and Quantum Engineering (YINQE), and by the Yale School of Engineering and Applied Science (SEAS) cleanroom.
This research was supported by the US Army Research Office (Grants No. W911NF-14-1-0563, No. W911NF-18-1-0020, No. W911NF-18-1-0212, and No. W911NF-14-1-0011, No. W911NF-16-1-0349) and the US Air Force Office of Scientific Research (Grant No. FA9550-15-1-0029).

\nocite{*}

\bibliographystyle{apsrev4-1}
\bibliography{Cav_atten_bib}

\begin{thebibliography}{41}%
\makeatletter
\providecommand \@ifxundefined [1]{%
 \@ifx{#1\undefined}
}%
\providecommand \@ifnum [1]{%
 \ifnum #1\expandafter \@firstoftwo
 \else \expandafter \@secondoftwo
 \fi
}%
\providecommand \@ifx [1]{%
 \ifx #1\expandafter \@firstoftwo
 \else \expandafter \@secondoftwo
 \fi
}%
\providecommand \natexlab [1]{#1}%
\providecommand \enquote  [1]{``#1''}%
\providecommand \bibnamefont  [1]{#1}%
\providecommand \bibfnamefont [1]{#1}%
\providecommand \citenamefont [1]{#1}%
\providecommand \href@noop [0]{\@secondoftwo}%
\providecommand \href [0]{\begingroup \@sanitize@url \@href}%
\providecommand \@href[1]{\@@startlink{#1}\@@href}%
\providecommand \@@href[1]{\endgroup#1\@@endlink}%
\providecommand \@sanitize@url [0]{\catcode `\\12\catcode `\$12\catcode
  `\&12\catcode `\#12\catcode `\^12\catcode `\_12\catcode `\%12\relax}%
\providecommand \@@startlink[1]{}%
\providecommand \@@endlink[0]{}%
\providecommand \url  [0]{\begingroup\@sanitize@url \@url }%
\providecommand \@url [1]{\endgroup\@href {#1}{\urlprefix }}%
\providecommand \urlprefix  [0]{URL }%
\providecommand \Eprint [0]{\href }%
\providecommand \doibase [0]{http://dx.doi.org/}%
\providecommand \selectlanguage [0]{\@gobble}%
\providecommand \bibinfo  [0]{\@secondoftwo}%
\providecommand \bibfield  [0]{\@secondoftwo}%
\providecommand \translation [1]{[#1]}%
\providecommand \BibitemOpen [0]{}%
\providecommand \bibitemStop [0]{}%
\providecommand \bibitemNoStop [0]{.\EOS\space}%
\providecommand \EOS [0]{\spacefactor3000\relax}%
\providecommand \BibitemShut  [1]{\csname bibitem#1\endcsname}%
\let\auto@bib@innerbib\@empty
\bibitem [{\citenamefont {Devoret}\ and\ \citenamefont
  {Schoelkopf}(2013)}]{Devoret13}%
  \BibitemOpen
  \bibfield  {author} {\bibinfo {author} {\bibfnamefont {M.~H.}\ \bibnamefont
  {Devoret}}\ and\ \bibinfo {author} {\bibfnamefont {R.~J.}\ \bibnamefont
  {Schoelkopf}},\ }\href {\doibase 10.1126/science.1231930} {\bibfield
  {journal} {\bibinfo  {journal} {Science}\ }\textbf {\bibinfo {volume}
  {339}},\ \bibinfo {pages} {1169} (\bibinfo {year} {2013})}\BibitemShut
  {NoStop}%
\bibitem [{\citenamefont {Vion}\ \emph {et~al.}(2002)\citenamefont {Vion},
  \citenamefont {Aassime}, \citenamefont {Cottet}, \citenamefont {Joyez},
  \citenamefont {Pothier}, \citenamefont {Urbina}, \citenamefont {Esteve},\
  and\ \citenamefont {Devoret}}]{Vion02}%
  \BibitemOpen
  \bibfield  {author} {\bibinfo {author} {\bibfnamefont {D.}~\bibnamefont
  {Vion}}, \bibinfo {author} {\bibfnamefont {A.}~\bibnamefont {Aassime}},
  \bibinfo {author} {\bibfnamefont {A.}~\bibnamefont {Cottet}}, \bibinfo
  {author} {\bibfnamefont {P.}~\bibnamefont {Joyez}}, \bibinfo {author}
  {\bibfnamefont {H.}~\bibnamefont {Pothier}}, \bibinfo {author} {\bibfnamefont
  {C.}~\bibnamefont {Urbina}}, \bibinfo {author} {\bibfnamefont
  {D.}~\bibnamefont {Esteve}}, \ and\ \bibinfo {author} {\bibfnamefont {M.~H.}\
  \bibnamefont {Devoret}},\ }\href {\doibase 10.1126/science.1069372}
  {\bibfield  {journal} {\bibinfo  {journal} {Science}\ }\textbf {\bibinfo
  {volume} {296}},\ \bibinfo {pages} {886} (\bibinfo {year}
  {2002})}\BibitemShut {NoStop}%
\bibitem [{\citenamefont {Cottet}(2002)}]{Cottet02}%
  \BibitemOpen
  \bibfield  {author} {\bibinfo {author} {\bibfnamefont {A.}~\bibnamefont
  {Cottet}},\ }\emph {\bibinfo {title} {Implementation of a quantum bit in a
  superconducting circuit}},\ \href
  {http://www.phys.ens.fr/~cottet/ACottetThesis.pdf} {\bibinfo {type} {{PhD}
  dissertation}},\ \bibinfo  {school} {Universite Paris VI} (\bibinfo {year}
  {2002})\BibitemShut {NoStop}%
\bibitem [{\citenamefont {Koch}\ \emph {et~al.}(2007)\citenamefont {Koch},
  \citenamefont {Yu}, \citenamefont {Gambetta}, \citenamefont {Houck},
  \citenamefont {Schuster}, \citenamefont {Majer}, \citenamefont {Blais},
  \citenamefont {Devoret}, \citenamefont {Girvin},\ and\ \citenamefont
  {Schoelkopf}}]{Koch07}%
  \BibitemOpen
  \bibfield  {author} {\bibinfo {author} {\bibfnamefont {J.}~\bibnamefont
  {Koch}}, \bibinfo {author} {\bibfnamefont {T.~M.}\ \bibnamefont {Yu}},
  \bibinfo {author} {\bibfnamefont {J.}~\bibnamefont {Gambetta}}, \bibinfo
  {author} {\bibfnamefont {A.~A.}\ \bibnamefont {Houck}}, \bibinfo {author}
  {\bibfnamefont {D.~I.}\ \bibnamefont {Schuster}}, \bibinfo {author}
  {\bibfnamefont {J.}~\bibnamefont {Majer}}, \bibinfo {author} {\bibfnamefont
  {A.}~\bibnamefont {Blais}}, \bibinfo {author} {\bibfnamefont {M.~H.}\
  \bibnamefont {Devoret}}, \bibinfo {author} {\bibfnamefont {S.~M.}\
  \bibnamefont {Girvin}}, \ and\ \bibinfo {author} {\bibfnamefont {R.~J.}\
  \bibnamefont {Schoelkopf}},\ }\href {\doibase 10.1103/PhysRevA.76.042319}
  {\bibfield  {journal} {\bibinfo  {journal} {Phys. Rev. A}\ }\textbf {\bibinfo
  {volume} {76}},\ \bibinfo {pages} {042319} (\bibinfo {year}
  {2007})}\BibitemShut {NoStop}%
\bibitem [{\citenamefont {Schreier}\ \emph {et~al.}(2008)\citenamefont
  {Schreier}, \citenamefont {Houck}, \citenamefont {Koch}, \citenamefont
  {Schuster}, \citenamefont {Johnson}, \citenamefont {Chow}, \citenamefont
  {Gambetta}, \citenamefont {Majer}, \citenamefont {Frunzio}, \citenamefont
  {Devoret}, \citenamefont {Girvin},\ and\ \citenamefont
  {Schoelkopf}}]{Schreier08}%
  \BibitemOpen
  \bibfield  {author} {\bibinfo {author} {\bibfnamefont {J.~A.}\ \bibnamefont
  {Schreier}}, \bibinfo {author} {\bibfnamefont {A.~A.}\ \bibnamefont {Houck}},
  \bibinfo {author} {\bibfnamefont {J.}~\bibnamefont {Koch}}, \bibinfo {author}
  {\bibfnamefont {D.~I.}\ \bibnamefont {Schuster}}, \bibinfo {author}
  {\bibfnamefont {B.~R.}\ \bibnamefont {Johnson}}, \bibinfo {author}
  {\bibfnamefont {J.~M.}\ \bibnamefont {Chow}}, \bibinfo {author}
  {\bibfnamefont {J.~M.}\ \bibnamefont {Gambetta}}, \bibinfo {author}
  {\bibfnamefont {J.}~\bibnamefont {Majer}}, \bibinfo {author} {\bibfnamefont
  {L.}~\bibnamefont {Frunzio}}, \bibinfo {author} {\bibfnamefont {M.~H.}\
  \bibnamefont {Devoret}}, \bibinfo {author} {\bibfnamefont {S.~M.}\
  \bibnamefont {Girvin}}, \ and\ \bibinfo {author} {\bibfnamefont {R.~J.}\
  \bibnamefont {Schoelkopf}},\ }\href {\doibase 10.1103/PhysRevB.77.180502}
  {\bibfield  {journal} {\bibinfo  {journal} {Phys. Rev. B}\ }\textbf {\bibinfo
  {volume} {77}},\ \bibinfo {pages} {180502} (\bibinfo {year}
  {2008})}\BibitemShut {NoStop}%
\bibitem [{\citenamefont {You}\ \emph {et~al.}(2007)\citenamefont {You},
  \citenamefont {Hu}, \citenamefont {Ashhab},\ and\ \citenamefont
  {Nori}}]{You07}%
  \BibitemOpen
  \bibfield  {author} {\bibinfo {author} {\bibfnamefont {J.~Q.}\ \bibnamefont
  {You}}, \bibinfo {author} {\bibfnamefont {X.}~\bibnamefont {Hu}}, \bibinfo
  {author} {\bibfnamefont {S.}~\bibnamefont {Ashhab}}, \ and\ \bibinfo {author}
  {\bibfnamefont {F.}~\bibnamefont {Nori}},\ }\href {\doibase
  10.1103/PhysRevB.75.140515} {\bibfield  {journal} {\bibinfo  {journal} {Phys.
  Rev. B}\ }\textbf {\bibinfo {volume} {75}},\ \bibinfo {pages} {140515}
  (\bibinfo {year} {2007})}\BibitemShut {NoStop}%
\bibitem [{\citenamefont {Steffen}\ \emph {et~al.}(2010)\citenamefont
  {Steffen}, \citenamefont {Kumar}, \citenamefont {DiVincenzo}, \citenamefont
  {Rozen}, \citenamefont {Keefe}, \citenamefont {Rothwell},\ and\ \citenamefont
  {Ketchen}}]{Steffen10}%
  \BibitemOpen
  \bibfield  {author} {\bibinfo {author} {\bibfnamefont {M.}~\bibnamefont
  {Steffen}}, \bibinfo {author} {\bibfnamefont {S.}~\bibnamefont {Kumar}},
  \bibinfo {author} {\bibfnamefont {D.~P.}\ \bibnamefont {DiVincenzo}},
  \bibinfo {author} {\bibfnamefont {J.~R.}\ \bibnamefont {Rozen}}, \bibinfo
  {author} {\bibfnamefont {G.~A.}\ \bibnamefont {Keefe}}, \bibinfo {author}
  {\bibfnamefont {M.~B.}\ \bibnamefont {Rothwell}}, \ and\ \bibinfo {author}
  {\bibfnamefont {M.~B.}\ \bibnamefont {Ketchen}},\ }\href {\doibase
  10.1103/PhysRevLett.105.100502} {\bibfield  {journal} {\bibinfo  {journal}
  {Phys. Rev. Lett.}\ }\textbf {\bibinfo {volume} {105}},\ \bibinfo {pages}
  {100502} (\bibinfo {year} {2010})}\BibitemShut {NoStop}%
\bibitem [{\citenamefont {Manucharyan}\ \emph {et~al.}(2009)\citenamefont
  {Manucharyan}, \citenamefont {Koch}, \citenamefont {Glazman},\ and\
  \citenamefont {Devoret}}]{Manucharyan09}%
  \BibitemOpen
  \bibfield  {author} {\bibinfo {author} {\bibfnamefont {V.~E.}\ \bibnamefont
  {Manucharyan}}, \bibinfo {author} {\bibfnamefont {J.}~\bibnamefont {Koch}},
  \bibinfo {author} {\bibfnamefont {L.~I.}\ \bibnamefont {Glazman}}, \ and\
  \bibinfo {author} {\bibfnamefont {M.~H.}\ \bibnamefont {Devoret}},\ }\href
  {\doibase 10.1126/science.1175552} {\bibfield  {journal} {\bibinfo  {journal}
  {Science}\ }\textbf {\bibinfo {volume} {326}},\ \bibinfo {pages} {113}
  (\bibinfo {year} {2009})}\BibitemShut {NoStop}%
\bibitem [{\citenamefont {Blais}\ \emph {et~al.}(2004)\citenamefont {Blais},
  \citenamefont {Huang}, \citenamefont {Wallraff}, \citenamefont {Girvin},\
  and\ \citenamefont {Schoelkopf}}]{Blais04}%
  \BibitemOpen
  \bibfield  {author} {\bibinfo {author} {\bibfnamefont {A.}~\bibnamefont
  {Blais}}, \bibinfo {author} {\bibfnamefont {R.-S.}\ \bibnamefont {Huang}},
  \bibinfo {author} {\bibfnamefont {A.}~\bibnamefont {Wallraff}}, \bibinfo
  {author} {\bibfnamefont {S.~M.}\ \bibnamefont {Girvin}}, \ and\ \bibinfo
  {author} {\bibfnamefont {R.~J.}\ \bibnamefont {Schoelkopf}},\ }\href
  {\doibase 10.1103/PhysRevA.69.062320} {\bibfield  {journal} {\bibinfo
  {journal} {Phys. Rev. A}\ }\textbf {\bibinfo {volume} {69}},\ \bibinfo
  {pages} {062320} (\bibinfo {year} {2004})}\BibitemShut {NoStop}%
\bibitem [{\citenamefont {Wallraff}\ \emph {et~al.}(2004)\citenamefont
  {Wallraff}, \citenamefont {Schuster}, \citenamefont {Blais}, \citenamefont
  {Frunzio}, \citenamefont {Huang}, \citenamefont {Majer}, \citenamefont
  {Kumar}, \citenamefont {Girvin},\ and\ \citenamefont
  {Schoelkopf}}]{Wallraff04}%
  \BibitemOpen
  \bibfield  {author} {\bibinfo {author} {\bibfnamefont {A.}~\bibnamefont
  {Wallraff}}, \bibinfo {author} {\bibfnamefont {D.~I.}\ \bibnamefont
  {Schuster}}, \bibinfo {author} {\bibfnamefont {A.}~\bibnamefont {Blais}},
  \bibinfo {author} {\bibfnamefont {L.}~\bibnamefont {Frunzio}}, \bibinfo
  {author} {\bibfnamefont {R.-S.}\ \bibnamefont {Huang}}, \bibinfo {author}
  {\bibfnamefont {J.}~\bibnamefont {Majer}}, \bibinfo {author} {\bibfnamefont
  {S.}~\bibnamefont {Kumar}}, \bibinfo {author} {\bibfnamefont {S.~M.}\
  \bibnamefont {Girvin}}, \ and\ \bibinfo {author} {\bibfnamefont {R.~J.}\
  \bibnamefont {Schoelkopf}},\ }\href {http://dx.doi.org/10.1038/nature02851}
  {\bibfield  {journal} {\bibinfo  {journal} {Nature}\ }\textbf {\bibinfo
  {volume} {431}},\ \bibinfo {pages} {162} (\bibinfo {year}
  {2004})}\BibitemShut {NoStop}%
\bibitem [{\citenamefont {Esteve}\ \emph {et~al.}(1986)\citenamefont {Esteve},
  \citenamefont {Devoret},\ and\ \citenamefont {Martinis}}]{Esteve86}%
  \BibitemOpen
  \bibfield  {author} {\bibinfo {author} {\bibfnamefont {D.}~\bibnamefont
  {Esteve}}, \bibinfo {author} {\bibfnamefont {M.~H.}\ \bibnamefont {Devoret}},
  \ and\ \bibinfo {author} {\bibfnamefont {J.~M.}\ \bibnamefont {Martinis}},\
  }\href {\doibase 10.1103/PhysRevB.34.158} {\bibfield  {journal} {\bibinfo
  {journal} {Phys. Rev. B}\ }\textbf {\bibinfo {volume} {34}},\ \bibinfo
  {pages} {158} (\bibinfo {year} {1986})}\BibitemShut {NoStop}%
\bibitem [{\citenamefont {Houck}\ \emph {et~al.}(2008)\citenamefont {Houck},
  \citenamefont {Schreier}, \citenamefont {Johnson}, \citenamefont {Chow},
  \citenamefont {Koch}, \citenamefont {Gambetta}, \citenamefont {Schuster},
  \citenamefont {Frunzio}, \citenamefont {Devoret}, \citenamefont {Girvin},\
  and\ \citenamefont {Schoelkopf}}]{Houck08}%
  \BibitemOpen
  \bibfield  {author} {\bibinfo {author} {\bibfnamefont {A.~A.}\ \bibnamefont
  {Houck}}, \bibinfo {author} {\bibfnamefont {J.~A.}\ \bibnamefont {Schreier}},
  \bibinfo {author} {\bibfnamefont {B.~R.}\ \bibnamefont {Johnson}}, \bibinfo
  {author} {\bibfnamefont {J.~M.}\ \bibnamefont {Chow}}, \bibinfo {author}
  {\bibfnamefont {J.}~\bibnamefont {Koch}}, \bibinfo {author} {\bibfnamefont
  {J.~M.}\ \bibnamefont {Gambetta}}, \bibinfo {author} {\bibfnamefont {D.~I.}\
  \bibnamefont {Schuster}}, \bibinfo {author} {\bibfnamefont {L.}~\bibnamefont
  {Frunzio}}, \bibinfo {author} {\bibfnamefont {M.~H.}\ \bibnamefont
  {Devoret}}, \bibinfo {author} {\bibfnamefont {S.~M.}\ \bibnamefont {Girvin}},
  \ and\ \bibinfo {author} {\bibfnamefont {R.~J.}\ \bibnamefont {Schoelkopf}},\
  }\href {\doibase 10.1103/PhysRevLett.101.080502} {\bibfield  {journal}
  {\bibinfo  {journal} {Phys. Rev. Lett.}\ }\textbf {\bibinfo {volume} {101}},\
  \bibinfo {pages} {080502} (\bibinfo {year} {2008})}\BibitemShut {NoStop}%
\bibitem [{\citenamefont {Paik}\ \emph {et~al.}(2011)\citenamefont {Paik},
  \citenamefont {Schuster}, \citenamefont {Bishop}, \citenamefont {Kirchmair},
  \citenamefont {Catelani}, \citenamefont {Sears}, \citenamefont {Johnson},
  \citenamefont {Reagor}, \citenamefont {Frunzio}, \citenamefont {Glazman},
  \citenamefont {Girvin}, \citenamefont {Devoret},\ and\ \citenamefont
  {Schoelkopf}}]{Paik11}%
  \BibitemOpen
  \bibfield  {author} {\bibinfo {author} {\bibfnamefont {H.}~\bibnamefont
  {Paik}}, \bibinfo {author} {\bibfnamefont {D.~I.}\ \bibnamefont {Schuster}},
  \bibinfo {author} {\bibfnamefont {L.~S.}\ \bibnamefont {Bishop}}, \bibinfo
  {author} {\bibfnamefont {G.}~\bibnamefont {Kirchmair}}, \bibinfo {author}
  {\bibfnamefont {G.}~\bibnamefont {Catelani}}, \bibinfo {author}
  {\bibfnamefont {A.~P.}\ \bibnamefont {Sears}}, \bibinfo {author}
  {\bibfnamefont {B.~R.}\ \bibnamefont {Johnson}}, \bibinfo {author}
  {\bibfnamefont {M.~J.}\ \bibnamefont {Reagor}}, \bibinfo {author}
  {\bibfnamefont {L.}~\bibnamefont {Frunzio}}, \bibinfo {author} {\bibfnamefont
  {L.~I.}\ \bibnamefont {Glazman}}, \bibinfo {author} {\bibfnamefont {S.~M.}\
  \bibnamefont {Girvin}}, \bibinfo {author} {\bibfnamefont {M.~H.}\
  \bibnamefont {Devoret}}, \ and\ \bibinfo {author} {\bibfnamefont {R.~J.}\
  \bibnamefont {Schoelkopf}},\ }\href {\doibase 10.1103/PhysRevLett.107.240501}
  {\bibfield  {journal} {\bibinfo  {journal} {Phys. Rev. Lett.}\ }\textbf
  {\bibinfo {volume} {107}},\ \bibinfo {pages} {240501} (\bibinfo {year}
  {2011})}\BibitemShut {NoStop}%
\bibitem [{\citenamefont {Rist{\`e}}\ \emph {et~al.}(2013)\citenamefont
  {Rist{\`e}}, \citenamefont {Bultink}, \citenamefont {Tiggelman},
  \citenamefont {Schouten}, \citenamefont {Lehnert},\ and\ \citenamefont
  {DiCarlo}}]{Riste13}%
  \BibitemOpen
  \bibfield  {author} {\bibinfo {author} {\bibfnamefont {D.}~\bibnamefont
  {Rist{\`e}}}, \bibinfo {author} {\bibfnamefont {C.~C.}\ \bibnamefont
  {Bultink}}, \bibinfo {author} {\bibfnamefont {M.~J.}\ \bibnamefont
  {Tiggelman}}, \bibinfo {author} {\bibfnamefont {R.~N.}\ \bibnamefont
  {Schouten}}, \bibinfo {author} {\bibfnamefont {K.~W.}\ \bibnamefont
  {Lehnert}}, \ and\ \bibinfo {author} {\bibfnamefont {L.}~\bibnamefont
  {DiCarlo}},\ }\href {http://dx.doi.org/10.1038/ncomms2936} {\bibfield
  {journal} {\bibinfo  {journal} {Nat. Commun.}\ }\textbf {\bibinfo {volume}
  {4}},\ \bibinfo {pages} {1913} (\bibinfo {year} {2013})}\BibitemShut
  {NoStop}%
\bibitem [{\citenamefont {Dial}\ \emph {et~al.}(2016)\citenamefont {Dial},
  \citenamefont {McClure}, \citenamefont {Poletto}, \citenamefont {Keefe},
  \citenamefont {Rothwell}, \citenamefont {Gambetta}, \citenamefont {Abraham},
  \citenamefont {Chow},\ and\ \citenamefont {Steffen}}]{Dial16}%
  \BibitemOpen
  \bibfield  {author} {\bibinfo {author} {\bibfnamefont {O.}~\bibnamefont
  {Dial}}, \bibinfo {author} {\bibfnamefont {D.~T.}\ \bibnamefont {McClure}},
  \bibinfo {author} {\bibfnamefont {S.}~\bibnamefont {Poletto}}, \bibinfo
  {author} {\bibfnamefont {G.~A.}\ \bibnamefont {Keefe}}, \bibinfo {author}
  {\bibfnamefont {M.~B.}\ \bibnamefont {Rothwell}}, \bibinfo {author}
  {\bibfnamefont {J.~M.}\ \bibnamefont {Gambetta}}, \bibinfo {author}
  {\bibfnamefont {D.~W.}\ \bibnamefont {Abraham}}, \bibinfo {author}
  {\bibfnamefont {J.~M.}\ \bibnamefont {Chow}}, \ and\ \bibinfo {author}
  {\bibfnamefont {M.}~\bibnamefont {Steffen}},\ }\href
  {http://stacks.iop.org/0953-2048/29/i=4/a=044001} {\bibfield  {journal}
  {\bibinfo  {journal} {Supercond. Sci. Technol.}\ }\textbf {\bibinfo {volume}
  {29}},\ \bibinfo {pages} {044001} (\bibinfo {year} {2016})}\BibitemShut
  {NoStop}%
\bibitem [{\citenamefont {Narla}\ \emph {et~al.}(2016)\citenamefont {Narla},
  \citenamefont {Shankar}, \citenamefont {Hatridge}, \citenamefont {Leghtas},
  \citenamefont {Sliwa}, \citenamefont {Zalys-Geller}, \citenamefont
  {Mundhada}, \citenamefont {Pfaff}, \citenamefont {Frunzio}, \citenamefont
  {Schoelkopf},\ and\ \citenamefont {Devoret}}]{Narla16}%
  \BibitemOpen
  \bibfield  {author} {\bibinfo {author} {\bibfnamefont {A.}~\bibnamefont
  {Narla}}, \bibinfo {author} {\bibfnamefont {S.}~\bibnamefont {Shankar}},
  \bibinfo {author} {\bibfnamefont {M.}~\bibnamefont {Hatridge}}, \bibinfo
  {author} {\bibfnamefont {Z.}~\bibnamefont {Leghtas}}, \bibinfo {author}
  {\bibfnamefont {K.~M.}\ \bibnamefont {Sliwa}}, \bibinfo {author}
  {\bibfnamefont {E.}~\bibnamefont {Zalys-Geller}}, \bibinfo {author}
  {\bibfnamefont {S.~O.}\ \bibnamefont {Mundhada}}, \bibinfo {author}
  {\bibfnamefont {W.}~\bibnamefont {Pfaff}}, \bibinfo {author} {\bibfnamefont
  {L.}~\bibnamefont {Frunzio}}, \bibinfo {author} {\bibfnamefont {R.~J.}\
  \bibnamefont {Schoelkopf}}, \ and\ \bibinfo {author} {\bibfnamefont {M.~H.}\
  \bibnamefont {Devoret}},\ }\href {\doibase 10.1103/PhysRevX.6.031036}
  {\bibfield  {journal} {\bibinfo  {journal} {Phys. Rev. X}\ }\textbf {\bibinfo
  {volume} {6}},\ \bibinfo {pages} {031036} (\bibinfo {year}
  {2016})}\BibitemShut {NoStop}%
\bibitem [{\citenamefont {Minev}\ \emph {et~al.}(2018)\citenamefont {Minev},
  \citenamefont {Mundhada}, \citenamefont {Shankar}, \citenamefont {Reinhold},
  \citenamefont {Gutierrez-Jauregui}, \citenamefont {Schoelkopf}, \citenamefont
  {Mirrahimi}, \citenamefont {Carmichael},\ and\ \citenamefont
  {Devoret}}]{Minev18}%
  \BibitemOpen
  \bibfield  {author} {\bibinfo {author} {\bibfnamefont {Z.~K.}\ \bibnamefont
  {Minev}}, \bibinfo {author} {\bibfnamefont {S.~O.}\ \bibnamefont {Mundhada}},
  \bibinfo {author} {\bibfnamefont {S.}~\bibnamefont {Shankar}}, \bibinfo
  {author} {\bibfnamefont {P.}~\bibnamefont {Reinhold}}, \bibinfo {author}
  {\bibfnamefont {R.}~\bibnamefont {Gutierrez-Jauregui}}, \bibinfo {author}
  {\bibfnamefont {R.~J.}\ \bibnamefont {Schoelkopf}}, \bibinfo {author}
  {\bibfnamefont {M.}~\bibnamefont {Mirrahimi}}, \bibinfo {author}
  {\bibfnamefont {H.~J.}\ \bibnamefont {Carmichael}}, \ and\ \bibinfo {author}
  {\bibfnamefont {M.~H.}\ \bibnamefont {Devoret}},\ }\href
  {https://arxiv.org/abs/1803.00545} {\  (\bibinfo {year} {2018})},\ \Eprint
  {http://arxiv.org/abs/1803.00545} {arXiv:1803.00545} \BibitemShut {NoStop}%
\bibitem [{\citenamefont {Bertet}\ \emph
  {et~al.}(2005{\natexlab{a}})\citenamefont {Bertet}, \citenamefont
  {Chiorescu}, \citenamefont {Burkard}, \citenamefont {Semba}, \citenamefont
  {Harmans}, \citenamefont {DiVincenzo},\ and\ \citenamefont
  {Mooij}}]{Bertet05a}%
  \BibitemOpen
  \bibfield  {author} {\bibinfo {author} {\bibfnamefont {P.}~\bibnamefont
  {Bertet}}, \bibinfo {author} {\bibfnamefont {I.}~\bibnamefont {Chiorescu}},
  \bibinfo {author} {\bibfnamefont {G.}~\bibnamefont {Burkard}}, \bibinfo
  {author} {\bibfnamefont {K.}~\bibnamefont {Semba}}, \bibinfo {author}
  {\bibfnamefont {C.~J. P.~M.}\ \bibnamefont {Harmans}}, \bibinfo {author}
  {\bibfnamefont {D.~P.}\ \bibnamefont {DiVincenzo}}, \ and\ \bibinfo {author}
  {\bibfnamefont {J.~E.}\ \bibnamefont {Mooij}},\ }\href {\doibase
  10.1103/PhysRevLett.95.257002} {\bibfield  {journal} {\bibinfo  {journal}
  {Phys. Rev. Lett.}\ }\textbf {\bibinfo {volume} {95}},\ \bibinfo {pages}
  {257002} (\bibinfo {year} {2005}{\natexlab{a}})}\BibitemShut {NoStop}%
\bibitem [{\citenamefont {Bertet}\ \emph
  {et~al.}(2005{\natexlab{b}})\citenamefont {Bertet}, \citenamefont
  {Chiorescu}, \citenamefont {Harmans},\ and\ \citenamefont
  {Mooij}}]{Bertet05b}%
  \BibitemOpen
  \bibfield  {author} {\bibinfo {author} {\bibfnamefont {P.}~\bibnamefont
  {Bertet}}, \bibinfo {author} {\bibfnamefont {I.}~\bibnamefont {Chiorescu}},
  \bibinfo {author} {\bibfnamefont {C.~J. P.~M.}\ \bibnamefont {Harmans}}, \
  and\ \bibinfo {author} {\bibfnamefont {J.~E.}\ \bibnamefont {Mooij}},\ }\href
  {https://arxiv.org/abs/cond-mat/0507290} {\  (\bibinfo {year}
  {2005}{\natexlab{b}})},\ \Eprint {http://arxiv.org/abs/cond-mat/0507290}
  {arXiv:cond-mat/0507290} \BibitemShut {NoStop}%
\bibitem [{\citenamefont {Clerk}\ and\ \citenamefont {Utami}(2007)}]{Clerk07}%
  \BibitemOpen
  \bibfield  {author} {\bibinfo {author} {\bibfnamefont {A.~A.}\ \bibnamefont
  {Clerk}}\ and\ \bibinfo {author} {\bibfnamefont {D.~W.}\ \bibnamefont
  {Utami}},\ }\href {\doibase 10.1103/PhysRevA.75.042302} {\bibfield  {journal}
  {\bibinfo  {journal} {Phys. Rev. A}\ }\textbf {\bibinfo {volume} {75}},\
  \bibinfo {pages} {042302} (\bibinfo {year} {2007})}\BibitemShut {NoStop}%
\bibitem [{\citenamefont {Sears}\ \emph {et~al.}(2012)\citenamefont {Sears},
  \citenamefont {Petrenko}, \citenamefont {Catelani}, \citenamefont {Sun},
  \citenamefont {Paik}, \citenamefont {Kirchmair}, \citenamefont {Frunzio},
  \citenamefont {Glazman}, \citenamefont {Girvin},\ and\ \citenamefont
  {Schoelkopf}}]{Sears12}%
  \BibitemOpen
  \bibfield  {author} {\bibinfo {author} {\bibfnamefont {A.~P.}\ \bibnamefont
  {Sears}}, \bibinfo {author} {\bibfnamefont {A.}~\bibnamefont {Petrenko}},
  \bibinfo {author} {\bibfnamefont {G.}~\bibnamefont {Catelani}}, \bibinfo
  {author} {\bibfnamefont {L.}~\bibnamefont {Sun}}, \bibinfo {author}
  {\bibfnamefont {H.}~\bibnamefont {Paik}}, \bibinfo {author} {\bibfnamefont
  {G.}~\bibnamefont {Kirchmair}}, \bibinfo {author} {\bibfnamefont
  {L.}~\bibnamefont {Frunzio}}, \bibinfo {author} {\bibfnamefont {L.~I.}\
  \bibnamefont {Glazman}}, \bibinfo {author} {\bibfnamefont {S.~M.}\
  \bibnamefont {Girvin}}, \ and\ \bibinfo {author} {\bibfnamefont {R.~J.}\
  \bibnamefont {Schoelkopf}},\ }\href {\doibase 10.1103/PhysRevB.86.180504}
  {\bibfield  {journal} {\bibinfo  {journal} {Phys. Rev. B}\ }\textbf {\bibinfo
  {volume} {86}},\ \bibinfo {pages} {180504} (\bibinfo {year}
  {2012})}\BibitemShut {NoStop}%
\bibitem [{\citenamefont {Yan}\ \emph {et~al.}(2016)\citenamefont {Yan},
  \citenamefont {Gustavsson}, \citenamefont {Kamal}, \citenamefont {Birenbaum},
  \citenamefont {Sears}, \citenamefont {Hover}, \citenamefont {Gudmundsen},
  \citenamefont {Rosenberg}, \citenamefont {Samach}, \citenamefont {Weber},
  \citenamefont {Yoder}, \citenamefont {Orlando}, \citenamefont {Clarke},
  \citenamefont {Kerman},\ and\ \citenamefont {Oliver}}]{Yan16}%
  \BibitemOpen
  \bibfield  {author} {\bibinfo {author} {\bibfnamefont {F.}~\bibnamefont
  {Yan}}, \bibinfo {author} {\bibfnamefont {S.}~\bibnamefont {Gustavsson}},
  \bibinfo {author} {\bibfnamefont {A.}~\bibnamefont {Kamal}}, \bibinfo
  {author} {\bibfnamefont {J.}~\bibnamefont {Birenbaum}}, \bibinfo {author}
  {\bibfnamefont {A.~P.}\ \bibnamefont {Sears}}, \bibinfo {author}
  {\bibfnamefont {D.}~\bibnamefont {Hover}}, \bibinfo {author} {\bibfnamefont
  {T.~J.}\ \bibnamefont {Gudmundsen}}, \bibinfo {author} {\bibfnamefont
  {D.}~\bibnamefont {Rosenberg}}, \bibinfo {author} {\bibfnamefont
  {G.}~\bibnamefont {Samach}}, \bibinfo {author} {\bibfnamefont
  {S.}~\bibnamefont {Weber}}, \bibinfo {author} {\bibfnamefont {J.~L.}\
  \bibnamefont {Yoder}}, \bibinfo {author} {\bibfnamefont {T.~P.}\ \bibnamefont
  {Orlando}}, \bibinfo {author} {\bibfnamefont {J.}~\bibnamefont {Clarke}},
  \bibinfo {author} {\bibfnamefont {A.~J.}\ \bibnamefont {Kerman}}, \ and\
  \bibinfo {author} {\bibfnamefont {W.~D.}\ \bibnamefont {Oliver}},\ }\href
  {http://dx.doi.org/10.1038/ncomms12964} {\bibfield  {journal} {\bibinfo
  {journal} {Nat. Commun.}\ }\textbf {\bibinfo {volume} {7}},\ \bibinfo {pages}
  {12964} (\bibinfo {year} {2016})}\BibitemShut {NoStop}%
\bibitem [{\citenamefont {Yeh}\ \emph {et~al.}(2017)\citenamefont {Yeh},
  \citenamefont {LeFebvre}, \citenamefont {Premaratne}, \citenamefont
  {Wellstood},\ and\ \citenamefont {Palmer}}]{Yeh17}%
  \BibitemOpen
  \bibfield  {author} {\bibinfo {author} {\bibfnamefont {J.-H.}\ \bibnamefont
  {Yeh}}, \bibinfo {author} {\bibfnamefont {J.}~\bibnamefont {LeFebvre}},
  \bibinfo {author} {\bibfnamefont {S.}~\bibnamefont {Premaratne}}, \bibinfo
  {author} {\bibfnamefont {F.~C.}\ \bibnamefont {Wellstood}}, \ and\ \bibinfo
  {author} {\bibfnamefont {B.~S.}\ \bibnamefont {Palmer}},\ }\href {\doibase
  10.1063/1.4984894} {\bibfield  {journal} {\bibinfo  {journal} {J. Appl.
  Phys.}\ }\textbf {\bibinfo {volume} {121}},\ \bibinfo {pages} {224501}
  (\bibinfo {year} {2017})}\BibitemShut {NoStop}%
\bibitem [{\citenamefont {Yan}\ \emph {et~al.}(2018)\citenamefont {Yan},
  \citenamefont {Campbell}, \citenamefont {Krantz}, \citenamefont {Kjaergaard},
  \citenamefont {Kim}, \citenamefont {Yoder}, \citenamefont {Hover},
  \citenamefont {Sears}, \citenamefont {Kerman}, \citenamefont {Orlando},
  \citenamefont {Gustavsson},\ and\ \citenamefont {Oliver}}]{Yan18}%
  \BibitemOpen
  \bibfield  {author} {\bibinfo {author} {\bibfnamefont {F.}~\bibnamefont
  {Yan}}, \bibinfo {author} {\bibfnamefont {D.}~\bibnamefont {Campbell}},
  \bibinfo {author} {\bibfnamefont {P.}~\bibnamefont {Krantz}}, \bibinfo
  {author} {\bibfnamefont {M.}~\bibnamefont {Kjaergaard}}, \bibinfo {author}
  {\bibfnamefont {D.}~\bibnamefont {Kim}}, \bibinfo {author} {\bibfnamefont
  {J.~L.}\ \bibnamefont {Yoder}}, \bibinfo {author} {\bibfnamefont
  {D.}~\bibnamefont {Hover}}, \bibinfo {author} {\bibfnamefont
  {A.}~\bibnamefont {Sears}}, \bibinfo {author} {\bibfnamefont {A.~J.}\
  \bibnamefont {Kerman}}, \bibinfo {author} {\bibfnamefont {T.~P.}\
  \bibnamefont {Orlando}}, \bibinfo {author} {\bibfnamefont {S.}~\bibnamefont
  {Gustavsson}}, \ and\ \bibinfo {author} {\bibfnamefont {W.~D.}\ \bibnamefont
  {Oliver}},\ }\href {https://arxiv.org/abs/1801.00467} {\  (\bibinfo {year}
  {2018})},\ \Eprint {http://arxiv.org/abs/1801.00467} {arXiv:1801.00467}
  \BibitemShut {NoStop}%
\bibitem [{\citenamefont {Gambetta}\ \emph {et~al.}(2006)\citenamefont
  {Gambetta}, \citenamefont {Blais}, \citenamefont {Schuster}, \citenamefont
  {Wallraff}, \citenamefont {Frunzio}, \citenamefont {Majer}, \citenamefont
  {Devoret}, \citenamefont {Girvin},\ and\ \citenamefont
  {Schoelkopf}}]{Gambetta06}%
  \BibitemOpen
  \bibfield  {author} {\bibinfo {author} {\bibfnamefont {J.}~\bibnamefont
  {Gambetta}}, \bibinfo {author} {\bibfnamefont {A.}~\bibnamefont {Blais}},
  \bibinfo {author} {\bibfnamefont {D.~I.}\ \bibnamefont {Schuster}}, \bibinfo
  {author} {\bibfnamefont {A.}~\bibnamefont {Wallraff}}, \bibinfo {author}
  {\bibfnamefont {L.}~\bibnamefont {Frunzio}}, \bibinfo {author} {\bibfnamefont
  {J.}~\bibnamefont {Majer}}, \bibinfo {author} {\bibfnamefont {M.~H.}\
  \bibnamefont {Devoret}}, \bibinfo {author} {\bibfnamefont {S.~M.}\
  \bibnamefont {Girvin}}, \ and\ \bibinfo {author} {\bibfnamefont {R.~J.}\
  \bibnamefont {Schoelkopf}},\ }\href {\doibase 10.1103/PhysRevA.74.042318}
  {\bibfield  {journal} {\bibinfo  {journal} {Phys. Rev. A}\ }\textbf {\bibinfo
  {volume} {74}},\ \bibinfo {pages} {042318} (\bibinfo {year}
  {2006})}\BibitemShut {NoStop}%
\bibitem [{\citenamefont {Suri}\ \emph {et~al.}(2013)\citenamefont {Suri},
  \citenamefont {Keane}, \citenamefont {Ruskov}, \citenamefont {Bishop},
  \citenamefont {Tahan}, \citenamefont {Novikov}, \citenamefont {Robinson},
  \citenamefont {Wellstood},\ and\ \citenamefont {Palmer}}]{Suri13}%
  \BibitemOpen
  \bibfield  {author} {\bibinfo {author} {\bibfnamefont {B.}~\bibnamefont
  {Suri}}, \bibinfo {author} {\bibfnamefont {Z.~K.}\ \bibnamefont {Keane}},
  \bibinfo {author} {\bibfnamefont {R.}~\bibnamefont {Ruskov}}, \bibinfo
  {author} {\bibfnamefont {L.~S.}\ \bibnamefont {Bishop}}, \bibinfo {author}
  {\bibfnamefont {C.}~\bibnamefont {Tahan}}, \bibinfo {author} {\bibfnamefont
  {S.}~\bibnamefont {Novikov}}, \bibinfo {author} {\bibfnamefont {J.~E.}\
  \bibnamefont {Robinson}}, \bibinfo {author} {\bibfnamefont {F.~C.}\
  \bibnamefont {Wellstood}}, \ and\ \bibinfo {author} {\bibfnamefont {B.~S.}\
  \bibnamefont {Palmer}},\ }\href {\doibase 10.1088/1367-2630/15/12/125007}
  {\bibfield  {journal} {\bibinfo  {journal} {New J. Phys.}\ }\textbf {\bibinfo
  {volume} {15}},\ \bibinfo {pages} {125007} (\bibinfo {year}
  {2013})}\BibitemShut {NoStop}%
\bibitem [{\citenamefont {Rigetti}\ \emph {et~al.}(2012)\citenamefont
  {Rigetti}, \citenamefont {Gambetta}, \citenamefont {Poletto}, \citenamefont
  {Plourde}, \citenamefont {Chow}, \citenamefont {C\'orcoles}, \citenamefont
  {Smolin}, \citenamefont {Merkel}, \citenamefont {Rozen}, \citenamefont
  {Keefe}, \citenamefont {Rothwell}, \citenamefont {Ketchen},\ and\
  \citenamefont {Steffen}}]{Rigetti12}%
  \BibitemOpen
  \bibfield  {author} {\bibinfo {author} {\bibfnamefont {C.}~\bibnamefont
  {Rigetti}}, \bibinfo {author} {\bibfnamefont {J.~M.}\ \bibnamefont
  {Gambetta}}, \bibinfo {author} {\bibfnamefont {S.}~\bibnamefont {Poletto}},
  \bibinfo {author} {\bibfnamefont {B.~L.~T.}\ \bibnamefont {Plourde}},
  \bibinfo {author} {\bibfnamefont {J.~M.}\ \bibnamefont {Chow}}, \bibinfo
  {author} {\bibfnamefont {A.~D.}\ \bibnamefont {C\'orcoles}}, \bibinfo
  {author} {\bibfnamefont {J.~A.}\ \bibnamefont {Smolin}}, \bibinfo {author}
  {\bibfnamefont {S.~T.}\ \bibnamefont {Merkel}}, \bibinfo {author}
  {\bibfnamefont {J.~R.}\ \bibnamefont {Rozen}}, \bibinfo {author}
  {\bibfnamefont {G.~A.}\ \bibnamefont {Keefe}}, \bibinfo {author}
  {\bibfnamefont {M.~B.}\ \bibnamefont {Rothwell}}, \bibinfo {author}
  {\bibfnamefont {M.~B.}\ \bibnamefont {Ketchen}}, \ and\ \bibinfo {author}
  {\bibfnamefont {M.}~\bibnamefont {Steffen}},\ }\href {\doibase
  10.1103/PhysRevB.86.100506} {\bibfield  {journal} {\bibinfo  {journal} {Phys.
  Rev. B}\ }\textbf {\bibinfo {volume} {86}},\ \bibinfo {pages} {100506}
  (\bibinfo {year} {2012})}\BibitemShut {NoStop}%
\bibitem [{\citenamefont {Goetz}\ \emph {et~al.}(2017)\citenamefont {Goetz},
  \citenamefont {Pogorzalek}, \citenamefont {Deppe}, \citenamefont {Fedorov},
  \citenamefont {Eder}, \citenamefont {Fischer}, \citenamefont {Wulschner},
  \citenamefont {Xie}, \citenamefont {Marx},\ and\ \citenamefont
  {Gross}}]{Goetz17}%
  \BibitemOpen
  \bibfield  {author} {\bibinfo {author} {\bibfnamefont {J.}~\bibnamefont
  {Goetz}}, \bibinfo {author} {\bibfnamefont {S.}~\bibnamefont {Pogorzalek}},
  \bibinfo {author} {\bibfnamefont {F.}~\bibnamefont {Deppe}}, \bibinfo
  {author} {\bibfnamefont {K.~G.}\ \bibnamefont {Fedorov}}, \bibinfo {author}
  {\bibfnamefont {P.}~\bibnamefont {Eder}}, \bibinfo {author} {\bibfnamefont
  {M.}~\bibnamefont {Fischer}}, \bibinfo {author} {\bibfnamefont
  {F.}~\bibnamefont {Wulschner}}, \bibinfo {author} {\bibfnamefont
  {E.}~\bibnamefont {Xie}}, \bibinfo {author} {\bibfnamefont {A.}~\bibnamefont
  {Marx}}, \ and\ \bibinfo {author} {\bibfnamefont {R.}~\bibnamefont {Gross}},\
  }\href {\doibase 10.1103/PhysRevLett.118.103602} {\bibfield  {journal}
  {\bibinfo  {journal} {Phys. Rev. Lett.}\ }\textbf {\bibinfo {volume} {118}},\
  \bibinfo {pages} {103602} (\bibinfo {year} {2017})}\BibitemShut {NoStop}%
\bibitem [{\citenamefont {Ashcroft}\ and\ \citenamefont
  {Mermin}(1976)}]{Ashcroft}%
  \BibitemOpen
  \bibfield  {author} {\bibinfo {author} {\bibfnamefont {N.~W.}\ \bibnamefont
  {Ashcroft}}\ and\ \bibinfo {author} {\bibfnamefont {N.}~\bibnamefont
  {Mermin}},\ }\href
  {https://www.cengage.com/c/solid-state-physics-1e-ashcroft/9780030839931}
  {\emph {\bibinfo {title} {Solid State Physics}}}\ (\bibinfo  {publisher}
  {Cengage},\ \bibinfo {address} {Boston},\ \bibinfo {year} {1976})\BibitemShut
  {NoStop}%
\bibitem [{\citenamefont {Roukes}\ \emph {et~al.}(1985)\citenamefont {Roukes},
  \citenamefont {Freeman}, \citenamefont {Germain}, \citenamefont
  {Richardson},\ and\ \citenamefont {Ketchen}}]{Roukes85}%
  \BibitemOpen
  \bibfield  {author} {\bibinfo {author} {\bibfnamefont {M.~L.}\ \bibnamefont
  {Roukes}}, \bibinfo {author} {\bibfnamefont {M.~R.}\ \bibnamefont {Freeman}},
  \bibinfo {author} {\bibfnamefont {R.~S.}\ \bibnamefont {Germain}}, \bibinfo
  {author} {\bibfnamefont {R.~C.}\ \bibnamefont {Richardson}}, \ and\ \bibinfo
  {author} {\bibfnamefont {M.~B.}\ \bibnamefont {Ketchen}},\ }\href {\doibase
  10.1103/PhysRevLett.55.422} {\bibfield  {journal} {\bibinfo  {journal} {Phys.
  Rev. Lett.}\ }\textbf {\bibinfo {volume} {55}},\ \bibinfo {pages} {422}
  (\bibinfo {year} {1985})}\BibitemShut {NoStop}%
\bibitem [{\citenamefont {Wellstood}\ \emph {et~al.}(1994)\citenamefont
  {Wellstood}, \citenamefont {Urbina},\ and\ \citenamefont
  {Clarke}}]{Wellstood94}%
  \BibitemOpen
  \bibfield  {author} {\bibinfo {author} {\bibfnamefont {F.~C.}\ \bibnamefont
  {Wellstood}}, \bibinfo {author} {\bibfnamefont {C.}~\bibnamefont {Urbina}}, \
  and\ \bibinfo {author} {\bibfnamefont {J.}~\bibnamefont {Clarke}},\ }\href
  {\doibase 10.1103/PhysRevB.49.5942} {\bibfield  {journal} {\bibinfo
  {journal} {Phys. Rev. B}\ }\textbf {\bibinfo {volume} {49}},\ \bibinfo
  {pages} {5942} (\bibinfo {year} {1994})}\BibitemShut {NoStop}%
\bibitem [{\citenamefont {Pozar}(2012)}]{Pozar}%
  \BibitemOpen
  \bibfield  {author} {\bibinfo {author} {\bibfnamefont {D.~M.}\ \bibnamefont
  {Pozar}},\ }\href
  {https://www.wiley.com/en-us/Microwave+Engineering\%2C+4th+Edition-p-9781118298138}
  {\emph {\bibinfo {title} {Microwave Engineering}}},\ \bibinfo {edition}
  {4th}\ ed.\ (\bibinfo  {publisher} {Wiley},\ \bibinfo {address} {Hobojen},\
  \bibinfo {year} {2012})\BibitemShut {NoStop}%
\bibitem [{\citenamefont {Ho}\ \emph {et~al.}(2004)\citenamefont {Ho},
  \citenamefont {Newman}, \citenamefont {Rahimifard},\ and\ \citenamefont
  {Allen}}]{Ho04}%
  \BibitemOpen
  \bibfield  {author} {\bibinfo {author} {\bibfnamefont {K.~H.}\ \bibnamefont
  {Ho}}, \bibinfo {author} {\bibfnamefont {S.~T.}\ \bibnamefont {Newman}},
  \bibinfo {author} {\bibfnamefont {S.}~\bibnamefont {Rahimifard}}, \ and\
  \bibinfo {author} {\bibfnamefont {R.~D.}\ \bibnamefont {Allen}},\ }\href
  {\doibase 10.1016/j.ijmachtools.2004.04.017} {\bibfield  {journal} {\bibinfo
  {journal} {Int. J. Mach. Tools Manuf.}\ }\textbf {\bibinfo {volume} {44}},\
  \bibinfo {pages} {1247} (\bibinfo {year} {2004})}\BibitemShut {NoStop}%
\bibitem [{EDM()}]{EDM}%
  \BibitemOpen
  \href@noop {} {}\bibinfo {note} {Wire EDM cutting was performed by Advanced
  Research Corporation (ARC)}\BibitemShut {NoStop}%
\bibitem [{\citenamefont {Brecht}\ \emph {et~al.}(2017)\citenamefont {Brecht},
  \citenamefont {Chu}, \citenamefont {Axline}, \citenamefont {Pfaff},
  \citenamefont {Blumoff}, \citenamefont {Chou}, \citenamefont {Krayzman},
  \citenamefont {Frunzio},\ and\ \citenamefont {Schoelkopf}}]{Brecht17}%
  \BibitemOpen
  \bibfield  {author} {\bibinfo {author} {\bibfnamefont {T.}~\bibnamefont
  {Brecht}}, \bibinfo {author} {\bibfnamefont {Y.}~\bibnamefont {Chu}},
  \bibinfo {author} {\bibfnamefont {C.}~\bibnamefont {Axline}}, \bibinfo
  {author} {\bibfnamefont {W.}~\bibnamefont {Pfaff}}, \bibinfo {author}
  {\bibfnamefont {J.~Z.}\ \bibnamefont {Blumoff}}, \bibinfo {author}
  {\bibfnamefont {K.}~\bibnamefont {Chou}}, \bibinfo {author} {\bibfnamefont
  {L.}~\bibnamefont {Krayzman}}, \bibinfo {author} {\bibfnamefont
  {L.}~\bibnamefont {Frunzio}}, \ and\ \bibinfo {author} {\bibfnamefont
  {R.~J.}\ \bibnamefont {Schoelkopf}},\ }\href {\doibase
  10.1103/PhysRevApplied.7.044018} {\bibfield  {journal} {\bibinfo  {journal}
  {Phys. Rev. Applied}\ }\textbf {\bibinfo {volume} {7}},\ \bibinfo {pages}
  {044018} (\bibinfo {year} {2017})}\BibitemShut {NoStop}%
\bibitem [{\citenamefont {Pippard}(1947)}]{Pippard47}%
  \BibitemOpen
  \bibfield  {author} {\bibinfo {author} {\bibfnamefont {A.~B.}\ \bibnamefont
  {Pippard}},\ }\href {\doibase 10.1098/rspa.1947.0122} {\bibfield  {journal}
  {\bibinfo  {journal} {Proc. Roy. Soc. A}\ }\textbf {\bibinfo {volume}
  {191}},\ \bibinfo {pages} {385} (\bibinfo {year} {1947})}\BibitemShut
  {NoStop}%
\bibitem [{\citenamefont {Chambers}(1950)}]{Chambers50}%
  \BibitemOpen
  \bibfield  {author} {\bibinfo {author} {\bibfnamefont {R.~G.}\ \bibnamefont
  {Chambers}},\ }\href {http://dx.doi.org/10.1038/165239b0} {\bibfield
  {journal} {\bibinfo  {journal} {Nature}\ }\textbf {\bibinfo {volume} {165}},\
  \bibinfo {pages} {239} (\bibinfo {year} {1950})}\BibitemShut {NoStop}%
\bibitem [{\citenamefont {Pippard}(1954)}]{Pippard54}%
  \BibitemOpen
  \bibfield  {author} {\bibinfo {author} {\bibfnamefont {A.~B.}\ \bibnamefont
  {Pippard}},\ }\enquote {\bibinfo {title} {Metallic conduction at high
  frequencies and low temperatures},}\ in\ \href {\doibase
  10.1016/S0065-2539(08)60130-4} {\emph {\bibinfo {booktitle} {Advances in
  Electronics and Electron Physics}}},\ Vol.~\bibinfo {volume} {6},\ \bibinfo
  {editor} {edited by\ \bibinfo {editor} {\bibfnamefont {L.}~\bibnamefont
  {Marton}}}\ (\bibinfo  {publisher} {Academic Press},\ \bibinfo {year}
  {1954})\ p.~\bibinfo {pages} {1}\BibitemShut {NoStop}%
\bibitem [{\citenamefont {Reed}\ \emph {et~al.}(2010)\citenamefont {Reed},
  \citenamefont {Johnson}, \citenamefont {Houck}, \citenamefont {DiCarlo},
  \citenamefont {Chow}, \citenamefont {Schuster}, \citenamefont {Frunzio},\
  and\ \citenamefont {Schoelkopf}}]{Reed10}%
  \BibitemOpen
  \bibfield  {author} {\bibinfo {author} {\bibfnamefont {M.~D.}\ \bibnamefont
  {Reed}}, \bibinfo {author} {\bibfnamefont {B.~R.}\ \bibnamefont {Johnson}},
  \bibinfo {author} {\bibfnamefont {A.~A.}\ \bibnamefont {Houck}}, \bibinfo
  {author} {\bibfnamefont {L.}~\bibnamefont {DiCarlo}}, \bibinfo {author}
  {\bibfnamefont {J.~M.}\ \bibnamefont {Chow}}, \bibinfo {author}
  {\bibfnamefont {D.~I.}\ \bibnamefont {Schuster}}, \bibinfo {author}
  {\bibfnamefont {L.}~\bibnamefont {Frunzio}}, \ and\ \bibinfo {author}
  {\bibfnamefont {R.~J.}\ \bibnamefont {Schoelkopf}},\ }\href {\doibase
  10.1063/1.3435463} {\bibfield  {journal} {\bibinfo  {journal} {Appl. Phys.
  Lett.}\ }\textbf {\bibinfo {volume} {96}},\ \bibinfo {pages} {203110}
  (\bibinfo {year} {2010})}\BibitemShut {NoStop}%
\bibitem [{\citenamefont {Narla}\ \emph {et~al.}(2014)\citenamefont {Narla},
  \citenamefont {Sliwa}, \citenamefont {Hatridge}, \citenamefont {Shankar},
  \citenamefont {Frunzio}, \citenamefont {Schoelkopf},\ and\ \citenamefont
  {Devoret}}]{Narla14}%
  \BibitemOpen
  \bibfield  {author} {\bibinfo {author} {\bibfnamefont {A.}~\bibnamefont
  {Narla}}, \bibinfo {author} {\bibfnamefont {K.~M.}\ \bibnamefont {Sliwa}},
  \bibinfo {author} {\bibfnamefont {M.}~\bibnamefont {Hatridge}}, \bibinfo
  {author} {\bibfnamefont {S.}~\bibnamefont {Shankar}}, \bibinfo {author}
  {\bibfnamefont {L.}~\bibnamefont {Frunzio}}, \bibinfo {author} {\bibfnamefont
  {R.~J.}\ \bibnamefont {Schoelkopf}}, \ and\ \bibinfo {author} {\bibfnamefont
  {M.~H.}\ \bibnamefont {Devoret}},\ }\href {\doibase 10.1063/1.4883373}
  {\bibfield  {journal} {\bibinfo  {journal} {Appl. Phys. Lett.}\ }\textbf
  {\bibinfo {volume} {104}},\ \bibinfo {pages} {232605} (\bibinfo {year}
  {2014})}\BibitemShut {NoStop}%
\bibitem [{\citenamefont {Frattini}\ \emph {et~al.}(2017)\citenamefont
  {Frattini}, \citenamefont {Vool}, \citenamefont {Shankar}, \citenamefont
  {Narla}, \citenamefont {Sliwa},\ and\ \citenamefont {Devoret}}]{Frattini17}%
  \BibitemOpen
  \bibfield  {author} {\bibinfo {author} {\bibfnamefont {N.~E.}\ \bibnamefont
  {Frattini}}, \bibinfo {author} {\bibfnamefont {U.}~\bibnamefont {Vool}},
  \bibinfo {author} {\bibfnamefont {S.}~\bibnamefont {Shankar}}, \bibinfo
  {author} {\bibfnamefont {A.}~\bibnamefont {Narla}}, \bibinfo {author}
  {\bibfnamefont {K.~M.}\ \bibnamefont {Sliwa}}, \ and\ \bibinfo {author}
  {\bibfnamefont {M.~H.}\ \bibnamefont {Devoret}},\ }\href {\doibase
  10.1063/1.4984142} {\bibfield  {journal} {\bibinfo  {journal} {Appl. Phys.
  Lett.}\ }\textbf {\bibinfo {volume} {110}},\ \bibinfo {pages} {222603}
  (\bibinfo {year} {2017})}\BibitemShut {NoStop}%
\end{thebibliography}%

\begin{table*} [htb]
\centering
\begin{tabular}{ c c|c c c c c c c}
 \hline
 \hline
  \multicolumn{2}{c|}{Attenuator} & \multicolumn{6}{c}{$T = 25\mK$} \\
  \hline
  Material & Gap ($\si{\um}$) & $f_\mathrm{ro}$ (GHz) & $\kappa_\mathrm{i}/2\pi$ (MHz) & $\kappa_\mathrm{c}/2\pi$ (MHz) & $\chi/2\pi$ (MHz) & $T_{2\mathrm{e}}/2T_1$ & $\bar{n}_\mathrm{th}$ & $T_\mathrm{eff}$ (mK)\\ 
 \hline
 None & n/a  & 7.573 & n/a & 16.5 & 1.5 & $0.72\pm 0.12$ & $\leq 4\times10^{-3}$ & $\leq 65$\\ 
 Al & 75 & 7.847 & n/a & 0.24  & 1.1 & $0.75 \pm 0.06$ & $\leq 1\times10^{-3}$ & $\leq 55$ \\ 
 Brass & 300 & 7.573 & 11.4 & 1.9 & 1.2 & $0.98\, ({+0.02}/{-0.08})$ &$\leq 4\times 10^{-4}$ & $\leq 46$ \\ 
 Cu & 75 & 7.857 & 7.1 & 0.9 & 1.1 & $1.00\, ({+0.00}/{-0.12})$ &$\leq 2\times 10^{-4}$ & $\leq 44$ \\ 
 \hline
 \hline
 \end{tabular}
\caption{Frequency $f_\mathrm{ro}$, rates $\kappa_\mathrm{i}$ and $\kappa_\mathrm{c}$ of the readout mode, dispersive shift $\chi$, $T_{2\mathrm{e}}/2T_1$, and the estimate upper bounds of $\bar{n}_\mathrm{th}$ and $T_\mathrm{eff}$ at $25 \mK$ for each experiment in Fig. \ref{fig4}. The $\kappa_\mathrm{i}$'s for the experiments without an attenuator and with an aluminum filter are below fitting errors.}
\label{tab3}
\end{table*}

\end{document}